\documentclass[twocolumn,prb,amsmath,amssymb,longbibliography,showpacs,preprintunmbers]{revtex4-1}

\usepackage{graphicx}
\usepackage{amsbsy}
\usepackage{amsmath}
\usepackage{color}

\usepackage{bm}					
\usepackage[mathlines]{lineno}	
\usepackage{amssymb}
\usepackage{dcolumn}
\usepackage{changes}
\usepackage{epstopdf}
\usepackage{natbib}
\usepackage{microtype}
\begin{document}

\title{Time-domain grating with a periodically driven qutrit}

\author{Yingying Han$^{1,2,3,\dagger}$, Xiao-Qing Luo$^{2,3,\dagger}$, Tie-Fu Li$^{4,2,*}$, Wenxian Zhang$^{1,*}$, Shuai-Peng Wang$^{2}$, J. S. Tsai$^{5,6}$, Franco Nori$^{5,7}$, and J. Q. You$^{3,2,}$}

\email{$^\dagger$These authors contributed equally to this work. \\
$^*$Correspondence and requests for materials should be addressed to T.F.L. (litf@tsinghua.edu.cn), W.Z. (wxzhang@whu.edu.cn), or J.Q.Y. (jqyou@zju.edu.cn).}

\affiliation{{\footnotesize $^{1}$School of Physics and Technology, Wuhan University, Wuhan, Hubei 430072, China}}
\affiliation{{\footnotesize $^{2}$Quantum Physics and Quantum Information Division, Beijing Computational Science Research Center, Beijing 100193, China}}
\affiliation{{\footnotesize $^{3}$Interdisciplinary Center of Quantum Information and Zhejiang Province Key Laboratory of Quantum Technology and Device, Department of Physics and State Key Laboratory of Modern Optical Instrumentation, Zhejiang University, Hangzhou 310027, China}}
\affiliation{{\footnotesize $^{4}$Institute of Microelectronics, Department of Microelectronics and Nanoelectronics and Tsinghua National Laboratory of Information Science and Technology, Tsinghua University, Beijing 100084, China}}
\affiliation{{\footnotesize $^{5}$Theoretical Quantum Physics Laboratory, RIKEN Cluster for Pioneering Research, Wako-shi, Saitama 351-0198, Japan}}
\affiliation{{\footnotesize $^{6}$Department of Physic, Tokyo University of Science, Kagurazaka, Shinjuku-ku, Tokyo 162-8601, Japan}}
\affiliation{{\footnotesize $^{7}$Department of Physics, The University of Michigan, Ann Arbor, Michigan 48109-1040, USA}}

\date{\today}

\begin{abstract}
Physical systems in the time domain may exhibit analogous phenomena in real space, such as time crystals, time-domain Fresnel lenses, and modulational interference in a qubit. Here we report the experimental realization of time-domain grating using a superconducting qutrit in periodically modulated probe and control fields via two schemes: Simultaneous modulation and complementary modulation. Both experimental and numerical results exhibit modulated Autler-Townes (AT) and modulation-induced diffraction (MID) effects. Theoretical results also confirm that the peak positions of the interference fringes of AT and MID effects are determined by the usual two-level relative phases, while the observed diffraction fringes, appearing only in the complementary modulation, are however related to the three-level relative phase. Further analysis indicates that such a single-atom time-domain diffraction originates from the correlation effect between the two time-domain gratings. Moreover, we find that the widths of the diffraction fringes are independent of the control-field power. Our results shed light on the experimental exploration of quantum coherence for modulated multi-level systems and may find promising applications in fast all-microwave switches and quantum-gate operations in the strong-driving regime.
\end{abstract}

\maketitle

\section{ Introduction}

A modulated physical system may exhibit quite different dynamics from its free evolution~\cite{PhysRevLett.98.013601}, such as the dynamical phase transition occurring in modulated systems~\cite{PhysRevA.43.1802, Garrahan2009First, A2006Environmentally}, decoherence suppression of a qubit dynamically decoupled from its environment~\cite{PhysRevLett.114.190502, PhysRevB.75.201302, PhysRevB.77.125336}, and Landau-Zener-Stuckelberg interference in modulated two and three-level systems~\cite{Shevchenko20101, PhysRevA.83.033614, Du2010Noise, PhysRevB.97.045405, nanolett.5b04356}. With the advancement of control technologies, the modulation of a physical system has great potential not only in designing novel quantum phenomena ~\cite{PhysRevLett.109.160401,Zhang:2017ci, Choi:2017ho, PhysRevLett.89.203003,Shammah2018Open} but also in deepening the understanding of the physics.

Interference phenomena, as a clear evidence for the quantum coherence of a system, appears in various modulated quantum two-level systems~\cite{Ono2018Quantum}, including semiconductor quantum dots~\cite{zrenner2002quantum-dot}, two-level atoms~\cite{PhysRevA.59.2269}, and superconducting quantum circuits~\cite{You2011Atomic}. Interference fringes of up to 20 photon transitions in a strongly driven superconducting flux qubit were reported and considered as a time-domain Mach-Zehnder interferometer~\cite{Oliver2005Mach}. By further applying a spin-echo pulse, noise-resistant geometric Mach-Zehnder interferometry was also demonstrated in a similar qubit system~\cite{PhysRevLett.112.027001}. However, as another concrete evidence of the quantum coherence, diffraction in the time domain has rarely been explored, though the optical diffraction in real space always occurs with the interference in a conventional Young's double-slit experiment~\cite{Zia2007Surface}.

In contrast to a qubit, a qutrit acts as a three-level artificial atom and provides more degrees of freedom, which make it attractive in designing new concept quantum devices and in demonstrating novel atomic and quantum-optics phenomena. Two famous examples are the Autler-Townes (AT) splitting and electromagnetically induced transparency~\cite{Peng2014EIT, PhysRevA.89.063822}, which exist in three-level quantum systems and have been demonstrated in, e.g., atomic gases~\cite{PhysRevLett.66.2593, PhysRevLett.74.666}, artificial atoms in superconducting quantum circuits~\cite{PhysRevLett.104.163601,PhysRevA.93.053838}, quantum dots~\cite{NPquantumdot}, optomechanics~\cite{NatureEIT}, and three-level meta-atoms in meta-materials~\cite{PhysRevLett.101.047401, PhysRevLett.101.253903}. Here we take the advantage of a superconducting qutrit to design a time-domain grating and explore the quantum coherence of this three-level system, with an emphasis on time-domain diffraction phenomena. The employed qutrit in our experiment is a tunable 3D transmon~\cite{PhysRevLett.107.240501,PhysRevLett.114.240501}, in which the Josephson junction is replaced by a symmetric SQUID~\cite{PhysRevA.93.053838}. The qutrit parameter can be tuned via this SQUID.  The initial state of the qutrit is prepared in its ground state by cooling it down to a cryogenic temperature of around 25 mK (see  Appendix~\ref{app:setup}). After the state evolution, the transmission coefficient of the readout signal is measured when the system reaches its steady state.

With this qutrit, we investigate the quantum coherence patterns under two periodic modulation schemes, i.e., using a simultaneous modulation and a complimentary modulation. Under the simultaneous modulation, the time-domain interference and the diffraction phenomena are tangled together, while under the complementary modulation the time-domain diffraction stands out from the interference. The observed quantum coherence patterns of the periodically modulated qutrit can be ascribed to the two-level relative phases and the three-level relative phase, respectively.

\section{Autler-Townes and modulated Autler-Townes effects }

\begin{figure*}[tbh]
 \centering
 \renewcommand{\figurename} {{\bf Fig}}
\includegraphics[width=6.5in]{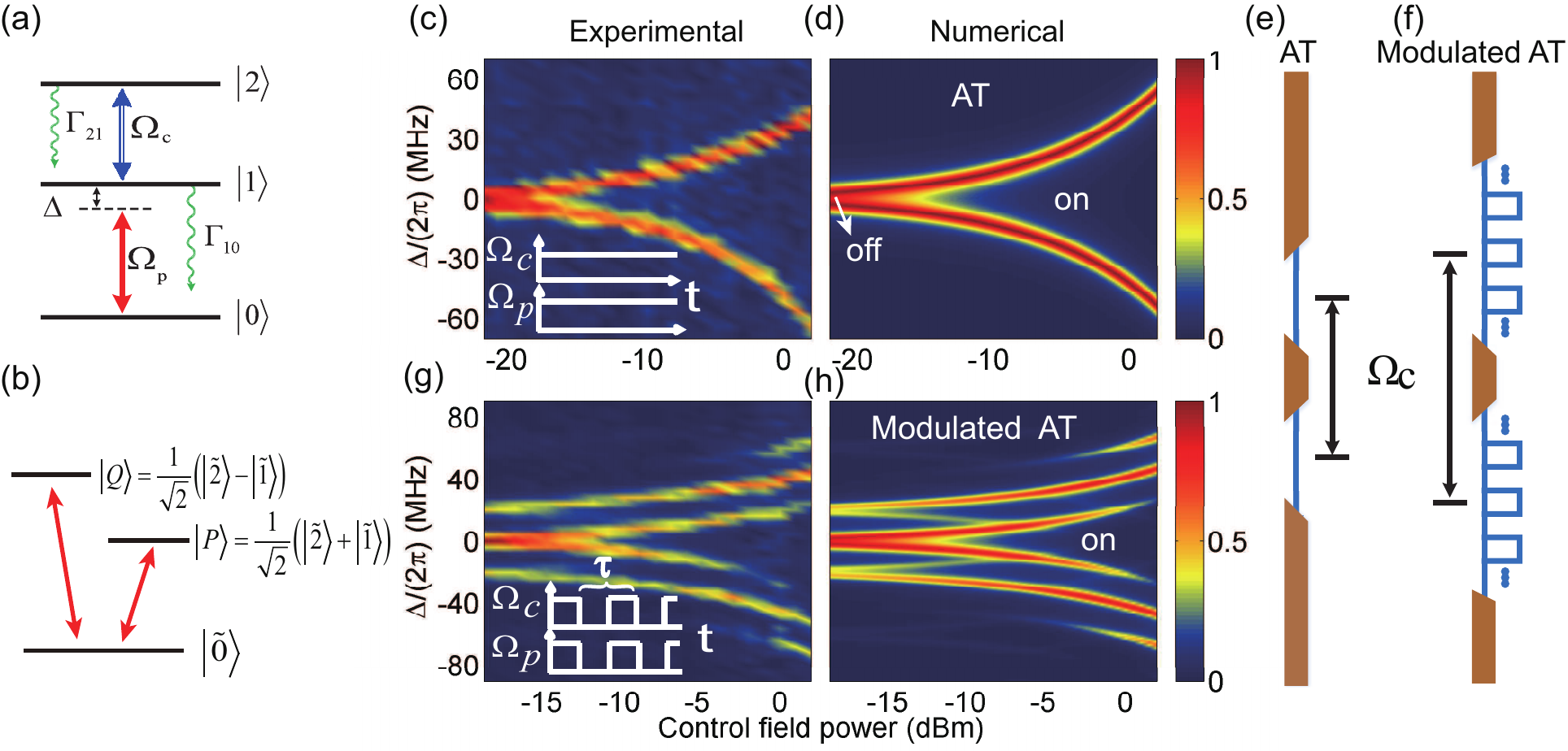}
\caption{\label{fig:coef}  Autler-Townes (AT) and modulated AT effects. (a) A qutrit is driven by a $\Delta$-detuned probe-field with strength $\Omega_p$ and a resonant control field with strength $\Omega_c$; $\Gamma_{21}$  ($\Gamma_{10}$) is the decay rate of the state $|2\rangle$ ($|1\rangle$). (b) Schematic of energy levels in the coupled basis. The strong control field splits the degenerate dressed states $|\tilde{1}\rangle$ and $|\tilde{2}\rangle$ into the AT doublet $|P\rangle$ and $|Q\rangle$. The coupling strength between the state $|P\rangle$ ($|Q\rangle$) and the ground state $|\tilde{0}\rangle$ is $-\Omega_p/\sqrt{2}$ ($\Omega_p/\sqrt{2}$). (c) and (d) Normalized experimental and numerical results of the AT effect with a probe-field power $-31$ dBm. (e) Analog of the AT effect to a light passing through two windows with spacing $\Omega_c$ in real space, which move in the opposite directions. (f) Analog of the modulated AT effect in real space, which is the same as in (e) except for a time-domain grating placed in each window, corresponding to the simultaneous modulations of the probe and control fields in the time-domain with a period $\tau$, as shown in the inset of (g). (g) and (h) Modulated AT effect under simultaneous modulations, with $\tau=50$ ns and a probe-field power $-31$ dBm.}
\end{figure*}

In our experiment, we apply two microwave fields to the qutrit, as shown in Fig.~\ref{fig:coef}(a). The resonantly coupled excited states $|1\rangle$ and $|2\rangle$ become degenerate in the dressed-state basis ($|\tilde{i}\rangle, i=0,1,2$) and further split into AT doublet due to the strong control field with a strength $\Omega_c$ [Fig.~\ref{fig:coef}(b)]~\cite{PhysRev.100.703, Peng2014EIT}. The AT doublet $|P\rangle$ and $|Q\rangle$ couple with the ground state $|\tilde{0}\rangle$ by the probe field with an effective strength $\Omega_p/\sqrt{2}$. In the limit of weak probe field, the respective phases of the three levels $|\tilde 0\rangle, |P\rangle$, and $|Q\rangle$ at time $t$ respectively are
\begin{eqnarray}\label{eq:phase}
\phi_0&=&-\Delta t/2,\nonumber\\
\phi_P&=&(\Delta-\Omega_c)t/2,\nonumber\\
\phi_Q&=&(\Delta+\Omega_c) t/2.
\end{eqnarray}
The experimental signal is proportional to $\rho_{PP}+\rho_{QQ}=\rho_{11}+\rho_{22}$ (see Appendix~\ref{app:nume}). The AT splitting effect in Figs~\ref{fig:coef}(c) and~\ref{fig:coef}(d) shows double fringes~\cite{PhysRev.100.703, PhysRevA.89.063822} with peaks satisfying the two-level relative phases $\theta_P =0$ and $\theta_Q=0$, where
\begin{eqnarray}\label{eq:phapq}
  \theta_P \equiv \phi_P-\phi_0&=&(2\Delta-\Omega_c)t/2,\nonumber\\
  \theta_Q \equiv \phi_Q-\phi_0&=&(2\Delta+\Omega_c)t/2.
\end{eqnarray}
These fringes at $\Delta = \pm \Omega_c/2$ manifest the resonant transitions $|P\rangle\leftrightarrow|\tilde{0}\rangle$ and $|Q\rangle\leftrightarrow|\tilde{0}\rangle$ for the steady state when $\theta_{P,Q}$ becomes zero (thus time-independent). Optically, the AT effect is analogous to a light passing through two windows in real space [Fig.~\ref{fig:coef}(e)]. Since here the widths of the windows are wide, only the effect of linear propagation appears. By increasing $\Omega_c$, the ``window" centers move in opposite directions with the same distance.

Next, we add modulations to explore the driven dynamics of the qutrit, in which the quantum coherence can demonstrate very different phenomena from a freely evolving three-level system. A straightforward scheme is to simultaneously modulate the control and probe fields in a square-wave with a period $\tau$ [cf. the inset in Fig.~\ref{fig:coef}(g)]. Such a modulation scheme is optically analogous to inserting time-domain slits with a finite slit width $\tau/2$ in each moving window [Fig.~\ref{fig:coef}(f)]. The effective number of slits in each window is estimated to be $1/(\Gamma \tau) \approx 4$, because $1/\Gamma \approx 200$ ns and $\tau = 50$ ns in our experiment (see Appendix~\ref{app:nume}).

Figures~\ref{fig:coef}(g) and~\ref{fig:coef}(h) show the modulated AT effect, which is similar to the conventional AT effect, except for the appearance of sidebands. The signal of the modulated AT pattern (see Appendix~\ref{app:grating} for details) can be interpreted as
\begin{eqnarray}
    \label{eq:mat}
  \rho_{11}+\rho_{22}  &\propto& D\left({3\theta_P-\theta_Q\over 8}\right)G\left({\theta_P\over 4}\right) \nonumber\\
  &+& D\left({3\theta_Q-\theta_P\over 8}\right)G\left({\theta_Q\over 4}\right),
\end{eqnarray}
where $\theta_P=\Delta \tau -\Omega_c \tau/4$ and $\theta_Q=\Delta \tau +\Omega_c \tau/4$ in a period; while $D(x) = \sin^2 x/x^2$ and $G(x) = \sin^2(2Nx)/\sin^2(2x)$ are, respectively, the diffraction function and the interference function of an $N$-slit grating in the time domain. Equation~(\ref{eq:mat}) shows that the modulated AT effect is the superposition of signals from two separate gratings moving in opposite directions. The widths of the upper branches in Figs.~\ref{fig:coef}(g) and~\ref{fig:coef}(h) are determined roughly by $|3\theta_P-\theta_Q| < 8\pi$, and the peak positions by $\theta_P=2n\pi$, with $n$ an integer. These peaks stem from the steady-state solution of the two-level relative phase under the periodic modulation, which should obey the time-domain Bloch theorem. Similarly, for the lower branches, the branch widths are determined by $|3\theta_Q-\theta_P| < 8\pi$, and the peak positions by $\theta_Q=2n\pi$.

\section{Time-domain modulation-induced diffraction effects}

\begin{figure*}[thb]
\centering
\renewcommand{\figurename} {{\bf Fig}}
\includegraphics[width=6.5in]{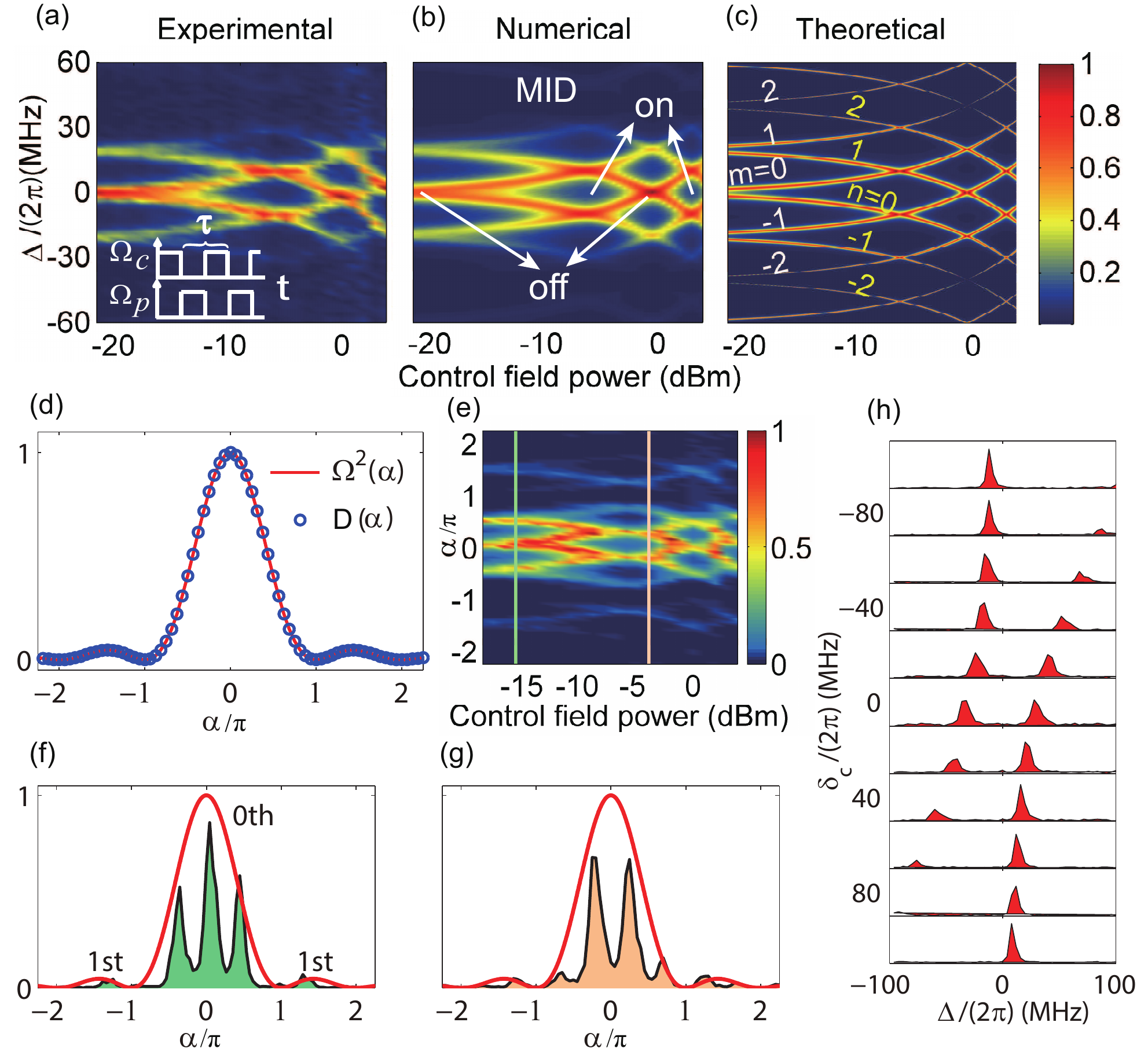}
\caption{\label{fig:peri} Time-domain modulation-induced diffraction effects. (a)-(c) Comparison of the experimentally-measured normalized transmission coefficient with numerical simulations and theoretical calculations. (a) Experimental results under the complementarily modulated probe and control fields with $\tau=50$ ns and a probe-field power $-31$ dBm, (b) Numerical simulations which include the damping rate $\Gamma_d$ of the qutrit resulting from its coupling to the environment, and (c) theoretical results using Floquet theory without including $\Gamma_d$ (i.e., obtained using Eq.~(\ref{eq:gvvf}) by only setting $\Gamma = 2\Omega(\alpha)$, see also Appendix~\ref{sec:gvv1}). The time-domain MID effect and power-broadening-free features are observed. (d) Comparison of the normalized $\Omega^2(\alpha)$ (red solid curve) with the diffraction function $D(\alpha)$ (blue circles). (e) Same as in (a) except for a probe-field power of $-20$ dBm. Note that the first-order diffractions become visible. (f) and (g) Experimental interference and the corresponding time-domain diffraction function (red curve) at the control-field powers  $-16$ dBm (i.e., the vertical green line in (e)) and $-4$ dBm (i.e., the vertical pink line in (e)). (h) Shift of the measured inteference peak versus the probe-field detuning $\Delta$ and the control-field detuning $\delta_{c}$ at the control-field power $-3$ dBm and the probe-field power $-33$ dBm.}
\end{figure*}

By designing a complementary modulation where the control and probe fields are complementarily varied in a square-wave [cf, the inset in Fig.~\ref{fig:peri}(a)], we explore more complex dynamics of the qutrit. Such a modulation scheme is very different from the modulated AT. Actually, as we show later, there is no direct optical analog of the gratings for the complementary modulation. The experimental results are presented in Fig.~\ref{fig:peri}(a).

For comparisons, numerical simulations and  theoretical predictions are also presented, respectively, in Figs.~\ref{fig:peri}(b) and~\ref{fig:peri}(c), where the damping rate $\Gamma_d$ of the qutrit resulting from its coupling to the environment is taken into account in the numerical simulations (Appendix~\ref{app:nume}),  but not included in the Floquet-theory calculations (Appendices~\ref{app:floquet} and~\ref{sec:gvv1}). We find a good agreement among these results, including the peak positions, peak widths, and interference patterns. In contrast to the AT and modulated AT effects, the modulation-induced diffraction (MID) results display a clear spectral concentration, i.e., the profile maximum of the interference pattern always lies at zero detuning ($\Delta = 0$) and becomes independent of the control-field power, clearly indicating that the signal is beyond the two-level relative phases.

Therefore, we introduce a three-level relative phase
\begin{equation}\label{eq:threep}
\theta_3=\theta_P+\theta_Q = \phi_P+\phi_Q-2\phi_0,
\end{equation}
which is the same as in a spinor Bose-Einstein condensate or in four-wave-mixing nonlinear optics~\cite{PhysRevA.72.013602, QNO, PhysRev.137.A801}. This relative phase $\theta_3$ appears if there is an interaction or nonlinear effect in the quantum system. We interpret the observed spectral concentration as a correlation effect between two time-domain gratings using this three-level relative phase.

For the complementarily-modulated strongly-driven qutrit, the time-averaged steady-state probability of the excited states, which is proportional to the measured transmission spectrum, is shown to be
\begin{eqnarray}
 \rho_{11}+\rho_{22}&\propto & D\left({\theta_3\over 8}\right)
\left[G\left({\theta_P\over 4}\right) + G\left({\theta_Q\over 4}\right)\right].
\label{eq:res1}
\end{eqnarray}
This theoretical result clearly explains the observed spectral concentration (Appendix~\ref{app:grating}). In particular, it can be seen that the diffraction function is independent of the control-field power, because $\theta_3 = 2\Delta \tau$ is independent of $\Omega_c$. The maximum of the coherence profile occurs at $\theta_3 = 0$ and the zero coherence at $\theta_3 = 8n\pi$ $(n\neq0)$ [see Fig.~\ref{fig:peri}(d)-(g)]. The explicit and sole dependence of the quantum diffraction on $\theta_3$ indicates that the MID effect is certainly beyond the two-level coherence, because all the three level's phases are involved. By comparing Eqs.~(\ref{eq:mat}) and~(\ref{eq:res1}), the correlation introduced in the complementary scheme actually mixes together the diffraction effects of the two time-domain gratings, which has no direct analog in optics.

A more rigorous calculation can be based on Floquet theory~\cite{Chu2004Beyond, Grifoni1998Driven}, where the periodic time-dependent Hamiltonian is transformed to a time-independent Floquet Hamiltonian with infinite dimensions. After adopting the generalized Van-Vleck nearly-degenerate perturbation theory~\cite{PhysRevA.32.377} (see Appendix~\ref{app:gvv}), we obtain

\begin{eqnarray}\label{eq:gvvf}
\rho_{11}+\rho_{22}
&=&\Omega^2(\alpha)\left[ \sum_{n=-\infty}^{\infty}\frac{1}{(\Delta-\Delta_n^P)^2 + \Gamma^2} \right.\nonumber\\
&+&\left. \sum_{m=-\infty}^{\infty} \frac{1}{(\Delta-\Delta_m^Q)^2 +\Gamma^2}\right]\label{eq:res2},
\end{eqnarray}
where
\begin{eqnarray}
  \Delta_n^P &=& -n\omega-\frac{\Omega_c}{4}, \nonumber\\
  \Delta_m^Q &=& -m\omega+\frac{\Omega_c}{4},
\end{eqnarray}
corresponding to the resonance peaks, and $\omega = 2\pi/\tau$ (the integers $n,m$ index the quasi-levels). Equation (\ref{eq:res2}) consists of a series of Lorentzians (i.e., peaks), each one with width $\Gamma$. Here the experimentally obtained damping rate $\Gamma$ is contributed by both the coupling term $\Omega(\alpha)$ due to the drive fields and the dissipation of the qutrit resulting from its coupling to the environment (see Appendix~\ref{sec:gvv1}). These Lorentzians have the same peak positions with the interference functions given by $\theta_P=2n\pi$ and $\theta_Q=2m\pi$ in Eq.~(\ref{eq:res1}). As shown in Fig.~\ref{fig:peri}(d), the envelope function $\Omega^2(\alpha)$ is the same as the diffraction function $D(\alpha)$. Different from Eq.~(\ref{eq:res1}), the inclusion of both the higher-order terms in $\Omega_p$ and the total damping rate $\Gamma^2 \gg 4\Omega^2(\alpha)$ causes each interference lineshape changing from $G(x)$ to a Lorentzian shape.

\section{High-order time-domain modulation-induced diffraction effects}
By increasing the probe-field power, we experimentally observe a high-order time-domain MID effect. The result for $\Omega_p = -20$ dBm and $\tau = 50$ ns is shown in Fig.~\ref{fig:peri}(e), where we demonstrate the first-order time-domain diffraction in addition to the central zeroth order. In Figs.~\ref{fig:peri}(f) and \ref{fig:peri}(g), we use the diffraction function $\Omega^2(\alpha)$ as the envelope function to fit the experimental results, showing that the profile of the interference agrees well with the diffraction function. Note that the measured interference peaks slightly shift rightward on the whole, but it does not occur when the control-field detuning $\delta_c$ becomes zero. This is because the circuit used is not an ideal three-level system. Actually, when the control-field power increases, the higher levels may affect the three-level ``atom", producing a nonzero detuning $\delta_c$. Indeed, as shown in Fig.~\ref{fig:peri}(h), the shifts of the measured interference \allowbreak{peaks} increase with the detuning $|\delta_c|$, confirming the above argument.

\section{Discussion and conclusion}
In conclusion, we have experimentally realized time-domain gratings with a superconducting qutrit under the simultaneous modulation and complementary modulation, respectively. We observe an interesting time-domain MID effect which distinguishes the collective three-level quantum coherence, involving the three-level relative phase $\theta_3$, from the usual two-level ones. This is instructive to explore the collective multi-level relative phase of the multi-level system. To quantify the quantum coherence of an $N$-level system, it seems more efficient to use a single collective $N$-level relative phase than $N(N-1)/2$ pairs of two-level relative phases.

The realization of MID in the time domain brings a power-broadening-free effect (i.e., the widths of the interference fringes are independent of the control-field power). This effect may have potential applications to the fast quantum gate operations in the strong driving regime. In general, a fast quantum gate (e.g., X-gate for a qubit) requires a strong driving field along the $x$-axis. The gate fidelity decreases as the driving-field intensity increases, due to the power broadening effect. However, with the MID, it is promising to design a quantum gate with its fidelity immune to the power broadening effect. Moreover, the time-domain MID effect demonstrated in the superconducting qutrit is a general phenomenon. It can be readily realized in other multi-level atomic and artificial atomic systems, including atomic gases, semiconductor quantum dots, nuclear spins, and ultracold quantum gases, because the modulation schemes require only square-waves and the effect is robust against experimental uncertainties.

The power-broadening-free effect may also find promising applications in fast microwave switches in the strong-driving regime. Compared to a standard microwave switch~\cite{Stern2017Strong} working in the AT regime [the ``on" and ``off" positions are labeled in Fig.~\ref{fig:coef}(d)], the qutrit under the complementary modulation with many closer ``on" and ``off" positions [Fig.~\ref{fig:peri}(b)] may have a faster switching rate. By combining the time-domain MID and modulated AT regimes, an even faster microwave phase switch may be developed if we solely change the modulation phase between the control and probe fields, which corresponds to changing between the complementary modulation [Fig.~\ref{fig:peri}(b)] and the simultaneous modulation [Fig.~\ref{fig:coef}(h)], but keeping the powers of the control and probe fields unchanged. These microwave switches could be used to switch either on or off the coupling between a cavity and an embedded quantum device in a future superconducting quantum apparatus.

\begin{acknowledgements}
This work is supported by the National Key Research and Development Program of China (Grant No.~2016YFA0301200), the NSFC (Grant Nos.~11574239, 91836101 and 11774022), the NSAF (Grant No.~U1530401), and Science Challenge Project (Grant No.~TZ2018003).
F.N. was partially supported by the MURI Center for Dynamic Magneto-Optics via the Air Force Office of Scientific Research (AFOSR) (FA9550-14-1-0040), Army Research Office (ARO) (Grant No. W911NF-18-1- 0358), Asian Office of Aerospace Research and Development (AOARD) (Grant No. FA2386-18-1-4045), Japan Science and Technology Agency (JST) (the Q-LEAP program, the ImPACT program and CREST Grant No. JP- MJCR1676), Japan Society for the Promotion of Science (JSPS) (JSPS-RFBR Grant No. 17-52-50023, JSPS-FWO Grant No. VS.059.18N), RIKEN-AIST Challenge Research Fund, and the John Templeton Foundation.
\end{acknowledgements}

\begin{appendix}

\section{Experimental setup}\label{app:setup}

\begin{figure}[tbh]
\centering
\includegraphics[width=3.2in]{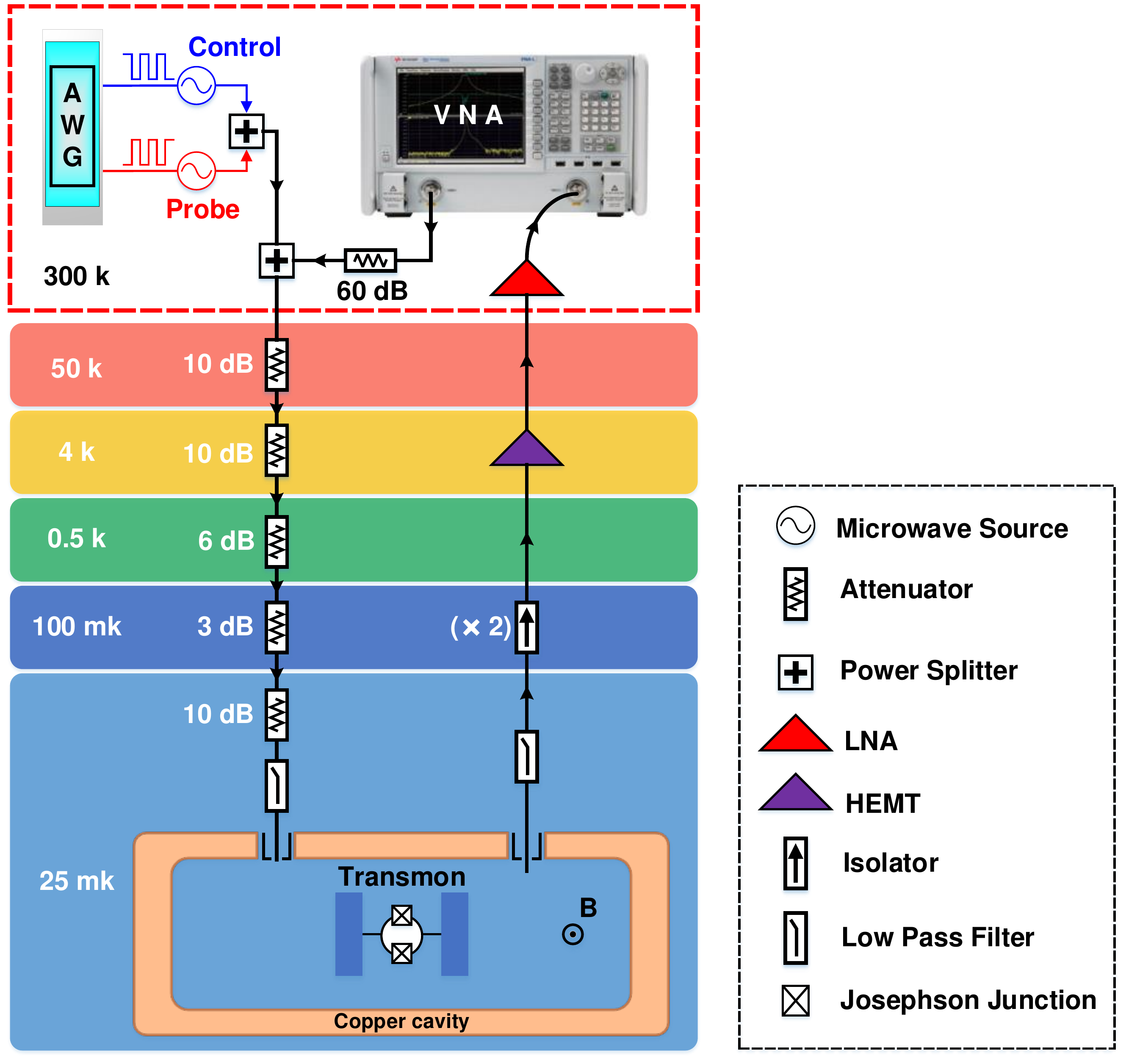}
\caption{\label{fig:exp}{Schematic diagram of the experimental setup.}}
\end{figure}

The measurements are performed in a BlueFors LD-400 dilution refrigerator at 25 mK. The sample used is a tunable three-dimensional (3D) transmon~\cite{PhysRevLett.107.240501,PhysRevLett.114.240501}, in which the Josephson junction is replaced by a symmetric SQUID with two identical Josephson junctions~\cite{PhysRevA.93.053838} (Fig.~\ref{fig:exp}) with two aluminium pads attached to the Josephson \allowbreak{junctions}. These two pads, together with the cavity, provide a large shunt capacitor to the SQUID. This shunt capacitor suppresses the charge noise to enhance the quantum coherence of the circuit, as in the C-shunt flux \allowbreak{qubits}~\cite{You-2007,Yan-2016} and 2D transmon. The circuit was fabricated on a $2\times7$~mm$^2$ silicon substrate patterned by electron-beam lithography, which was deposited using a standard double-angle evaporation process. The SQUID loop  size is $2\times 4$~$\mu$m$^2$ with two Al/AlO$_x$/Al junctions having an area of $140\times 150$~nm$^2$. Each shunting-capacitor Al pad has an area of  $250\times 500$~$\mu$m$^2$. The dimensions of the 3D cavity are $40\times 21\times 6$~mm$^3$, with its first eigenmode TE$_{101}$ having a resonant frequency of $\omega_{\mathrm{cavity}}/2\pi= 8.21996$ GHz and a photon decay rate of $\kappa/2\pi=1.46$ MHz. The vacuum Rabi coupling strength between the qubit and the first cavity mode is measured to be $g_0/\pi= 269.44$ MHz. To achieve a suitable Josephson coupling energy, we tune the magnetic flux in the SQUID loop. The obtained effective Josephson coupling energy is measured as $E_J/h=15.48$ GHz, and the charging energy of the 3D transmon is $E_C/h=337.32$ MHz. Here $|i\rangle$ ($i=0,1,2$) denote the lowest three eigenstates of the transmon, with transition frequencies $\omega_{01}/2\pi=6.12667$ GHz and $\omega_{12}/2\pi=5.78935$ GHz. This three-level system is used as a qutrit in our experiment.

In a microwave cavity-qubit system, the superconducting qubit is usually strongly coupled to the cavity. Provided that the bare frequency of the cavity is larger enough than the transition frequency of the qubit, the resonance frequency of the cavity shifts, depending on the state of the qubit. Therefore, the measurement of the transmission spectrum of the cavity can, in turn, be used to read out the state of the qubit~\cite{PhysRevLett.95.060501}. This dispersive readout technique is one of the quantum non-demolition measurement methods, which can also be applied to the superconducting qutrit~\cite{PhysRevA.93.053838}.

In our work, the transmission measurements are also performed in a dispersive regime. The readout signal tone is applied to the cavity at 8.22236 GHz, which corresponds to the resonant frequency of the 3D cavity when the population of the three-level system (qutrit) is initially in the ground state $|0\rangle$. As such, the measured transmission of the cavity, which is given by $T\equiv\rho_{00}T_0+\rho_{11}T_1+\rho_{22}T_2$, depends on the occupation probabilities of the qutrit, where $\rho_{00}+\rho_{11}+\rho_{22}=1$, and $T_0$, $T_1$, and $T_2$ represent the cavity transmission coefficients when the qutrit is in the state $|0\rangle$, $|1\rangle$, and $|2\rangle$, respectively. As shown in Ref.~\onlinecite{PhysRevA.93.053838}, the measured transmission coefficient of the cavity, which is measured using a vector network analyzer, is proportional to $\rho_{11}+\rho_{22}$ in our scheme. Thus, the quantum dynamics of the superconducting qutrit becomes effectively decoupled from the cavity mode in this dispersive regime.

The symmetric square-wave control and probe fields were generated by two microwave generators modulated by an arbitrary waveform generator (AWG), and were combined with the readout signal from the VNA at room temperature before being sent into the weakly coupled port of the 3D copper cavity. The output signal left from the 3D cavity via the strongly coupled port, passed through two isolators, and was then amplified by the high-electron-mobility transistor (HEMT) amplifier at 4K. The transmitted signal was further amplified by a low-noise amplifier (LNA) at room temperature, before returning back into the VNA.

\section{Numerical simulations}\label{app:nume}

For the modulated qutrit, the time-dependent Hamiltonian is
\begin{eqnarray}
\label{eq:h0}
  H(t)=\left(
  \begin{array}{c c c}
  -\frac{\Delta}{2} & -\Omega_p(t) &0  \\
  -\Omega_p(t) & \frac{\Delta}{2} & -\Omega_c(t) \\
  0 & -\Omega_c(t) & \frac{\Delta}{2}
  \end{array}
  \right).
\end{eqnarray}
 Equation (\ref{eq:h0}) is now written in terms of the dressed states $|\tilde 0\rangle, |\tilde 1\rangle$ and $|\tilde 2\rangle$, corresponding to the adoption of rotating frames. The Rabi coupling strengths $\Omega_{p,c}(t)$ are complementarily modulated by square waves with a $50\%$ duty cycle.

Numerical results are obtained by solving the following Lindblad master equation with experimentally obtained parameters~\cite{RevModPhys.77.633},
\begin{eqnarray}\label{eq:lme}
 \frac{\partial \rho (t)}{\partial t}&=& -i[H(t),\rho]+\frac{\Gamma_{10}}{2}(2\sigma_{01}\rho\sigma_{10}-\sigma_{11}\rho-\rho\sigma_{11}) \nonumber \\
 &+&\frac{\Gamma_{21}}{2}(2\sigma_{12}\rho\sigma_{21}-\sigma_{22}\rho-\rho\sigma_{22})\nonumber\\
 &+& \sum_{j=1,2} \gamma_{jj}(2\sigma_{jj}\rho\sigma_{jj}-\sigma_{jj}\rho-\rho\sigma_{jj}),
\end{eqnarray}
where the population damping rates are $\Gamma_{10}/2\pi=2.267$ MHz and $\Gamma_{21}/2\pi=4.534$ MHz. The dephasing rates are $\gamma_{11}/2\pi=\gamma_{22}/2\pi=0.9165$ MHz. The total damping rate is $\Gamma/2\pi=[(\Gamma_{10}+\Gamma_{21})/2+\gamma_{11}+\gamma_{22}]/2\pi=5.234$ MHz, which is larger than the typical value of $\Omega_n^P,~\Omega_m^Q$ ($\sim 1$ MHz). The atomic projection operator is $\sigma_{ij}=|i\rangle\langle j|$ ($i,j=0,1,2$). The conversion between the experimental microwave power and the corresponding Rabi frequency is $1$ dBm $=10 \log(1.38\times 10^{-4} \, \Omega_{c,p}^2)$, obtained by fitting the experimental data with a Lorentzian profile of the AT resonance. By setting the initial state in $\rho_{00}=1$, we calculate the population of the excited states after the system evolves.

By considering the time-independent modulated Hamiltonian~\cite{PhysRevA.93.053838}, we explicitly write down from Eq.~(\ref{eq:lme})
\begin{align}\label{eq:pme}
\frac{\partial \rho_{12}}{\partial t}&=-\Gamma\rho_{12}+\frac{i\Omega_{c}}{2}(\rho_{22}-\rho_{11}),\nonumber\\
\frac{\partial \rho_{22}}{\partial t}&=-\Gamma_{21}\rho_{22}+\frac{i\Omega_c}{2}(\rho_{12}-\rho_{21}).
\end{align}
Here we have neglected the ground state since $\Omega_p \ll \Omega_c$. In a steady state, from Eq.~(\ref{eq:pme}), we obtain
\begin{equation}\label{eq:rhor}
\rho_{22}=\frac{\Omega_c^2}{2\Gamma\Gamma_{21}+\Omega_c^2}\rho_{11}.
\end{equation}
Given typical experimental parameters $\Omega_{c}\gg\sqrt{2\Gamma\Gamma_{21}}=6.887$ MHz, we find $\rho_{22}\approx\rho_{11}$. This relation is also confirmed by our numerical simulations in the modulated scheme. Thus, the measured transmission spectrum $T$ of the cavity, proportional to the occupation probabilities of the two excited states of the three-level system (qutrit), is
\begin{equation}\label{eq:trans}
 T=A(\rho_{11}+\rho_{22}),
\end{equation}
where $A$ is a normalization constant. The results of the corresponding numerical simulations for the qutrit is shown in Fig.~2(b).

\section{Time-domain grating}\label{app:grating}

The square-wave modulation of the probe and control fields is equivalent to a time-domain grating. Here we present the analytical solution of the modulated AT and MID effects in the time domain by assuming the probe field power being a perturbation term (i.e. $\Omega_p^2/\Delta^2, \Omega_p^2/\Omega_c^2\rightarrow0$). The initial state of the qutrit is prepared in its ground state $|0\rangle$. The excited states' probability is
\begin{equation}\label{eq:rhopq}
 \rho_{11}+\rho_{22}=\rho_{PP}+\rho_{QQ},
\end{equation}
since
\begin{eqnarray}
  |P\rangle&=&(|\tilde 2\rangle + |\tilde 1\rangle)/\sqrt 2, \\
  |Q\rangle&=&(|\tilde 2\rangle -|\tilde 1\rangle)/\sqrt 2.
\end{eqnarray}
 The experimental signal is the average of $\rho_{11}+\rho_{22}$ in a modulation period, after the system reaches its steady state.

The general wave function has the form
\begin{equation}\label{eq:psipq}
|\psi(t)\rangle=c_P(t)|P\rangle+c_Q(t)|Q\rangle+c_{\tilde 0}(t)|0\rangle.
\end{equation}
For the simultaneous modulation, through an iterative approach, it is straightforward to find that the average value of $|c_P(t)|^2+|c_Q(t)|^2$ in a period is
\begin{eqnarray}\label{eq:arhopq}
 \rho_{PP}+\rho_{QQ}&=&D\left({3\theta_P-\theta_Q \over 8}\right)G\left({\theta_P\over 4}\right)\nonumber\\
 &+& D\left({3\theta_Q-\theta_P\over 8}\right)G\left({\theta_Q\over 4}\right).
\end{eqnarray}
Apparently, the signal is the product of the diffraction function and the interference function, which indicates that the experimental signal in the modulated AT case is exactly the same as two separate but oppositely moving gratings in optics. For the complementary modulation, however, the average value of $|c_Q(t)|^2+|c_Q(t)|^2$ in a period is
\begin{equation}\label{eq:brhopq}
\rho_{PP}+\rho_{QQ}=D\left({\theta_3\over 8}\right)\left[G\left({\theta_P\over 4}\right) + G\left({\theta_Q\over 4}\right)\right].
\end{equation}
The diffraction function may be interpreted as the mixing or the correlation effect between the two time-domain gratings.

\section{Floquet theory}\label{app:floquet}
A periodic time-dependent Hamiltonian can be  transformed to an equivalent time-independent infinite-dimensional Floquet matrix eigenvalue problem~\cite{Chu2004Beyond, Kohler2005Driven}. The Hamiltonian $H(t)$ has Fourier components of $\omega$,
\begin{eqnarray}
\label{eq:fhf}
  H(t) &=&\sum_{n=-n_c}^{n_c}H^{[n]} \exp(-in\omega t),
\end{eqnarray}
where $H^{[n]}$ are spanned by any orthogonal basis set. The cutoff is set as $n_c=50$ and $n_c=40$ for the results in Figs.~2(b) and~2(e), respectively. By employing the composite Floquet basis $|\alpha,n\rangle$ with $\alpha=0,1,2$, one obtains the infinite-dimensional Floquet matrix $H_{F}$, with the elements defined by
\begin{equation}\label{eq:nfm}
 \langle \alpha n|H_{F}|\beta m\rangle=H_{\alpha \beta}^{[n-m]} + n\hbar\omega\delta_{\alpha\beta}\delta_{n m}.
\end{equation}
The Floquet matrix is then diagonalized,
\begin{equation}\label{eq:fme}
  H_{F}|q_{\gamma l}\rangle=q_{\gamma l}|q_{\gamma l}\rangle,
\end{equation}
where $q_{\gamma l}$ is a quasi-level eigenvalue and $|q_{\gamma l}\rangle$ the corresponding eigenvector. Starting from the ground state $|\beta\rangle = |0\rangle$, the time-averaged probability $\rho_{\alpha\alpha}$ of the first excited state (with $\alpha = 1$) becomes
\begin{equation}\label{eq:thp}
  \rho_{\alpha\alpha}=\sum_{n}\sum_{\gamma l}|\langle\alpha n|q_{\gamma l}\rangle\langle q_{\gamma l}|\beta 0\rangle|^{2}.
\end{equation}

For the qutrit, we expand the square-wave modulation function $\Omega_p(t)$ and $\Omega_c(t)$ in Eq.~(\ref{eq:h0}) into many Fourier components

\begin{eqnarray}\label{eq:tran}
\Omega_p(t)&=&\frac{\Omega_{p}}{4}-\sum\limits_{n=1}^{\infty}(-1)^{n}\Omega_{pn}\cos(\omega_{n}t),\\
\Omega_c(t)&=&\frac{\Omega_{c}}{4}-\sum\limits_{n=1}^{\infty}(-1)^{n+1}\Omega_{cn}\cos(\omega_{n}t),
\end{eqnarray}
where
\begin{eqnarray}\label{eq:omegapn}
\Omega_{pn}&=&\frac{\Omega_{p}}{(2n-1)\pi},~~~\Omega_{cn}=\frac{\Omega_{c}}{(2n-1)\pi}, \nonumber\\
\omega_{n}&=&(2n-1)\omega,~~~~n=1,2,3\cdots. \nonumber
\end{eqnarray}
According to Floquet theory, the Floquet Hamiltonian $H_F$ has diagonally many Floquet matrix blocks
\begin{eqnarray}
\label{eq:trans}
 H_F^{[0]} &=& \left(
 \begin{array}{ccc}
  -\frac{\Delta}{2} &-\frac{\Omega_{p}}{4}&0\vspace{2mm}\\
  -\frac{\Omega_{p}}{4}& \frac{\Delta}{2}&-\frac{\Omega_{c}}{4}\vspace{2mm}\\
  0&-\frac{\Omega_{c}}{4}&\frac{\Delta}{2}
  \end{array}
 \right),
\end{eqnarray}

\begin{eqnarray}
 H_F^{[2n-1]} &=& H_F^{[-(2n-1)]} \nonumber \\
 &=&\left(
 \begin{array}{ccc}
  0 &\frac{(-1)^n\Omega_{pn}}{2}&0\vspace{2mm}\nonumber\\
  \frac{(-1)^n\Omega_{pn}}{2}& 0&\frac{(-1)^{n+1}\Omega_{cn}}{2}\vspace{2mm}\\
  0&\frac{(-1)^{n+1}\Omega_{cn}}{2}&0
  \end{array}\right),\nonumber\\
   H_F^{[2n]}&=&H_F^{[-2n]}=0. \nonumber
\end{eqnarray}

Note that in each nonvanishing Floquet matrix block, $\Omega_{pn}$ and $\Omega_{cn}$ have opposite sign, because the probe and control fields are modulated in complementary square-waves. We obtain the results shown in Fig.~2(c) by truncating the dimension of the Floquet matrix to $n_c=40$.

\section{Analytic solution under the nearly degenerate condition}\label{app:gvv}
\label{sec:gvv1}

To analytically solve the Floquet Hamiltonian for the modulated qutrit, we employ the generalized Van-Vleck (GVV) nearly-degenerate perturbation theory. The original Hamiltonian in Eq.~(\ref{eq:trans}) has large off-diagonal matrix elements ($\sim \Omega_c$) in the strongly driven regime. To utilize the GVV perturbation theory, we resort to a unitary transformation which transforms the Hamiltonian to the coupled space of dressed states $|\tilde0\rangle$, $|P\rangle=(|\tilde2\rangle+|\tilde1\rangle)/\sqrt{2}$, and $|Q\rangle=(|\tilde2\rangle-|\tilde1\rangle)/\sqrt{2}$, where the off-diagonal elements become small and proportional to $\Omega_p$. Then, we follow the nearly degenerated GVV theory to reduce the infinite-dimensional Floquet Hamiltonian to a $3\times3$ effective Hamiltonian which includes three nearly degenerate quasi-levels (we have neglected the higher-order correction),
\begin{equation}\label{eq:gvv}
 H_{\rm GVV}=\left(
 \begin{array}{c c c}
 -\frac{\Delta}{2}&\Omega_{n}^{P}&\Omega_{m}^{Q}\vspace{2mm}\\
 \Omega_{n}^{P}&\frac{\Delta}{2}-\frac{\Omega_{c}}{4}-n\omega & 0 \vspace{2mm}\\
 \Omega_{m}^{Q}& 0 &\frac{\Delta}{2}+\frac{\Omega_{c}}{4}-m\omega
 \end{array}
 \right),
\end{equation}
where
\begin{small}
\begin{align}
\label{qe:fmb}
 \Omega_n^{P}=&~\Omega_{p} \sum_{l=-\infty}^{\infty} \cdots \sum_{g=-\infty}^{\infty}\Big\{\frac{-1}{4\sqrt{2}}J_{\zeta}(B)
 +\sum_{j=1}^{\infty}A\big[J_{\zeta+(2n+1)}(B)   \notag \\
 &+J_{\zeta-(2n-1)}(B)\big]\Big\}J_{l}\left(\frac{-B}{9}\right)\cdots J_{g}\bigg(\frac{(-1)^{q-1} B}{(2q-1)^2 }\bigg),    \notag \\
 \Omega_n^{Q}=&~\Omega_{p} \sum_{l=-\infty}^{\infty} \cdots \sum_{g=-\infty}^{\infty}\Big\{\frac{1}{4\sqrt{2}}J_{\zeta}(-B)
 -\sum_{j=1}^{\infty}A\big[J_{\zeta+(2n+1)}(-B) \notag  \\
 &+J_{\zeta-(2n-1)}(-B)\big]\Big\}J_{l}\left(\frac{B}{9}\right)\cdots J_{g}\bigg(\frac{(-1)^{q} B}{(2q-1)^2 }\bigg)
 \end{align}
 \end{small}
\!\!\!with $\zeta=n-3l-\cdots-(2q-1)g$, $A=\frac{(-1)^j}{2\sqrt{2}(2j-1)\pi}$, and $B=\frac{-\Omega_{c}}{\omega\pi}$.
It is easy to numerically check that
\begin{equation}
(\Omega_n^P)^2 = (\Omega_{-n}^Q)^2.
\end{equation}
We notice that the normalized $\Omega_n^P$ is, in general, a function of $n$, $\Omega_c$, and $\omega$, i.e., $\Omega_n^P = \Omega (n,\Omega_c,\omega)$. We also find that the off-diagonal elements $(\Omega_n^P)^2$ are simply shifted for different $n$, following a relation
\begin{equation}\label{eq:omegamn}
\Omega^2(n,\Omega_c,\omega)=\Omega^2(0,\Omega_c/4+n\omega).
\end{equation}
Consequently, under the resonance condition $\Delta_n^P=-n\omega-\Omega_c/4$, and considering the even-function property of $(\Omega_0^P)^2$, we find a simple relation
\begin{eqnarray}
 (\Omega_n^P)^2=\Omega^2(\Delta_n^P),
\end{eqnarray}
similarly, $(\Omega_m^Q)^2=\Omega^2(\Delta_m^Q)$. Thus, we are able to define
\begin{equation}\label{eq:omegapq}
\Omega^2(\alpha) = \Omega^2(\Delta_n^P)=\Omega^2(\Delta_m^Q),
\end{equation}
with $\alpha=\Delta \tau/4$. Finally, the time-averaged steady-state probability of the excited states is calculated and presented in Eq.~(\ref{eq:res2}), where the total damping rate $\Gamma$ is included. The linewidths of the two kinds of resonances are the same and given by $\Gamma=\sqrt{\Gamma_d^2+4\Omega^2(\alpha)}$, where $\Gamma_d$ is the damping rate of the qutrit resulting from its coupling to the environment and the coupling term $\Omega(\alpha)$ is due to the effect of the drive fields. In Fig.~\ref{fig:peri}, the numerical simulations and theoretical calculations are compared with the experimental results, which indicate that $\Gamma_d^2\gg4\Omega^2(\alpha)$, i.e., $\Gamma\approx\Gamma_d$.

\end{appendix}

\end{document}